\documentclass[manuscript]{aastex}

\shorttitle{Line intensity enhancement in \ion{O}{6} }
\shortauthors{Keenan et al.}

\begin{document}


\title{Intensity enhancement of \ion{O}{6} ultraviolet emission lines in solar spectra due to opacity}


\author{F. P. Keenan} 
\affil{Astrophysics Research Centre, School of Mathematics and Physics, Queen's University Belfast, Belfast BT7 1NN, UK}
\email{f.keenan@qub.ac.uk}

\author{J. G. Doyle and M. S. Madjarska}
\affil{Armagh Observatory, College Hill, Armagh BT61 9DG, UK}

\author{S. J. Rose, L. A. Bowler\altaffilmark{1}, and J. Britton}
\affil{Physics Department, Imperial College, London SW7 2BZ, UK}

\author{L. McCrink}
\affil{Centre for Statistical Science and Operational Research, School of Mathematics and Physics, Queen's University Belfast, Belfast BT7 1NN, UK}

\and

\author{M. Mathioudakis}
\affil{Astrophysics Research Centre, School of Mathematics and Physics, Queen's University Belfast, Belfast BT7 1NN, UK}






\altaffiltext{1}{Current address: Computational Biology, Department of Computer Science, University of Oxford, Rex Richards Building, South Parks Road, Oxford OX1 3QU, UK}



\begin{abstract}
Opacity is a property of many plasmas, and it is normally expected that if an emission line in a plasma becomes optically thick, its intensity ratio to that of another transition that remains optically thin should decrease. However, radiative transfer calculations undertaken both by ourselves and others predict that under certain conditions the intensity ratio of an optically thick to thin line can show an increase over the optically thin value, indicating an enhancement in the former. These conditions include the geometry of the emitting plasma and its orientation to the observer. A similar effect can take place between lines of differing optical depth. Previous observational studies have focused on stellar point sources, and here we investigate the spatially-resolved solar atmosphere using measurements of the I(1032 \AA)/I(1038 \AA) intensity ratio of \ion{O}{6} in several regions obtained with the Solar Ultraviolet Measurements of Emitted Radiation (SUMER) instrument on board the Solar and Heliospheric Observatory (SoHO) satellite. We find several I(1032 \AA)/I(1038 \AA) ratios observed on the disk to be significantly larger than the optically thin value of 2.0, providing the first detection (to our knowledge) of intensity enhancement in the ratio 
arising from opacity effects in the solar atmosphere. Agreement between observation and theory is excellent, and confirms that the \ion{O}{6} emission originates 
from a slab-like geometry in the solar atmosphere, rather than from cylindrical structures. 
\end{abstract}


\keywords{Opacity --- Sun: transition region --- Sun: UV radiation}



\section{Introduction}

Opacity is a common property of many astrophysical and laboratory plasmas. In most circumstances, one would expect that opacity in an emission line would lead to a reduction in its intensity compared to the optically thin value. However, theoretical work 
by Bhatia and co-workers (Bhatia \& Kastner 1999; Bhatia \& Saba 2001; Kastner \& Bhatia 2001) using the escape factor method indicated that in certain circumstances the intensity of an emission line could be enhanced over its optically thin value due to the effects of opacity. This research did not explain how this (apparently counter-intuitive) result came about, which had to await the more sophisticated calculations of Kerr et al. (2004),  who determined the radiation transport in the plasma using the CRETIN code (Scott 2001). The CRETIN results provided the origin of the line enhancement effect, namely that the ion in an upper state of a transition can be pumped in the optically thick case by photons traversing the plasma at many different angles. As a consequence, the line intensity enhancement effect, and its apparent magnitude, is very dependent both on the geometry of the plasma and the orientation of the observer (i.e. the line-of-sight to the plasma by which it is viewed). Subsequently, Kerr et al. (2005) extended this work by using an analytical approach to consider several different geometries. They found that the detection of line intensity enhancement could, in theory, discriminate between different plasma geometries and the orientation of the observer. This would in principle provide a powerful new diagnostic for astrophysical sources, many of which are spatially unresolved. 

Observationally, searches have been undertaken for line intensity enhancements in stellar spectra, using the ratio of lines of differing optical depth. Rose et al. (2008) found some evidence for the effect in the I(15.01 \AA)/I(16.78 \AA) ratio of \ion{Fe}{17} for the active cool dwarf EV Lac, with a measured value of 2.50$\pm$0.50 from XMM-Newton satellite observations compared to a theoretical optically thin result of $\leq$\,1.93. More recently, Keenan et al. (2011) analysed Far Ultraviolet Spectroscopic Explorer (FUSE) satellite spectra of the active late-type stars $\epsilon$ Eri, II Peg and Prox Cen, and measured several I(1032 \AA)/I(1038 \AA) ratios of \ion{O}{6} that were larger (by up to 30\%) than the optically thin value of 2.0. 

Although we are confident that the above detections are secure, they are very limited in number, and additionally are restricted to spatially unresolved (distant stellar) objects. In the present paper we therefore extend the work to search for \ion{O}{6} line intensity enhancements in a spatially resolved source, namely the Sun, and also 
build on our previous theoretical research for \ion{O}{6} (Keenan et al. 2011) to calculate \ion{O}{6} models for cylindrical as well as slab and spherical geometries.

\section{Observations}


Observations were obtained with the Solar Ultraviolet Measurements of Emitted Radiation (SUMER) instrument 
(Wilhelm et al. 1995) on board the Solar and Heliospheric Observatory (SoHO) satellite on 1996 June 7, centered 
at xcen = 980, ycen = 309 using a 1.125\arcsec\ step and 10~s exposure time from 
11:02:29 UT to 12:12:36 UT. Only two spectral windows, each with 25 spectral pixels, were analysed. 
Standard SUMER data reduction procedures were used, including decompression and reversion, dead-time, 
local gain and flat-field corrections, a correction for geometrical distortion, plus radiometric 
calibration correcting for data taken on both the bare and Kbr parts of detector A. 

Intensities of the \ion{O}{6} 1032 \AA\ (1s$^{2}$2s $^{2}$S$_{1/2}$--1s$^{2}$2p $^{2}$P$_{3/2}$) and 1038 \AA\ 
(1s$^{2}$2s $^{2}$S$_{1/2}$--1s$^{2}$2p $^{2}$P$_{1/2}$) lines were measured for three regions within the 
SUMER dataset, designated A, B and C 
as shown in Figure 1. For each region, \ion{O}{6} line intensities were determined in 
a 7 pixel $\times$ 7 pixel area, then 3 pixels were moved in the horizontal direction and the measurements repeated. 
We checked the flux of each line using two techniques, firstly a summation under the line profile 
and secondly via a Gaussian fit, in both cases removing the underlying continuum and instrumental background. 
The two sets of line intensities agreed to better than 1\%, and the former 
were adopted. Resultant I(1032)/I(1038) line intensity ratios are plotted as a function of position for each region in Figure 1. 
 
\section{Theory}

The basic theory behind the enhancement of emission line intensities due to opacity, and its dependence on plasma geometry, was discussed in detail by Kerr et al. (2004, 2005). Subsequently, Keenan et al. (2011) extended this research to the specific case of two lines of \ion{O}{6}, namely 1032 and 1038 \AA. Hence we refer the reader to these papers for further information, and here only provide a brief overview of the key results and our extension to the research since the publication of Keenan et al. For the 1032 and 1038 \AA\ lines of \ion{O}{6}, where their electron impact excitation rates are predicted to differ by a factor of 2.0 (Aggarwal \&\ Keenan 2004), their intensity ratio is

\begin{equation}
\frac{I(1032)}{I(1038)} = 2 \frac{g_a(1032)g_b(1038)}{g_b(1032)g_a(1038)}
\end{equation}

where g$_a$(k) is the probability for photon escape in the line-of-sight of the observer for line k and g$_b$(k) is the angle-averaged escape probability. In the case of a spherical plasma, g$_a$(k) = g$_b$(k) by symmetry, and hence
I(1032)/I(1038) = 2.0 for the coronal steady-state plasma assumed in all our modelling. 
For an infinite plane slab

\begin{equation}
\frac{I(1032)}{I(1038)} = 2 \frac{\displaystyle\int_0^\infty \left (1-\exp\left[\frac{-2\tau_0}{\mu^\prime} e^{-x^2} \right] \right)dx}{\displaystyle\int_0^\infty \left (1-\exp\left[\frac{-\tau_0}{\mu^\prime} e^{-x^2} \right] \right)dx} \frac{\displaystyle\int_0^\infty \int_0^1 \mu \left (1-\exp\left[\frac{-\tau_0}{\mu} e^{-x^2} \right] \right) d\mu dx }{\displaystyle\int_0^\infty \int_0^1 \mu \left (1-\exp\left[\frac{-2\tau_0}{\mu} e^{-x^2} \right] \right) d\mu dx} 
\end{equation}

where $\tau_0$ is the optical depth of the 1038 \AA\ transition at line center, which has a pathlength that is the 
perpendicular thickness ({\em l}) of the plasma slab, and 
$\mu^\prime$ = $cos \theta$ where $\theta$ is the angle between the perpendicular to the slab and the line-of-sight
to the observer, as shown in Figure 2. 

We have now extended this work to derive the line intensity ratio for an infinite cylinder, which is

\begin{tiny}

\begin{equation}
\frac{I(1032)}{I(1038)} = 2 \frac{\displaystyle\int_0^1 \int_0^\infty \frac{\alpha}{\sqrt{1-\alpha^2}} \left (1-\exp\left[\frac{-2\tau_0\alpha}{\sqrt{1-\mu^{\prime 2}}} e^{-x^2} \right] \right) dxd\alpha}{\displaystyle\int_0^1 \int_0^\infty \frac{\alpha}{\sqrt{1-\alpha^2}} \left (1-\exp\left[\frac{-\tau_0\alpha}{\sqrt{1-\mu^{\prime 2}}} e^{-x^2} \right] \right) dxd\alpha} \frac{\displaystyle \int_0^1 \int_0^1 \int_0^\infty \alpha \frac{\sqrt{1-\beta^2}}{\sqrt{1-\alpha^2}} \left (1-\exp\left[\frac{-\tau_0\alpha}{\sqrt{1-\beta^2}} e^{-x^2} \right] \right) dxd\alpha d\beta}{\displaystyle \int_0^1 \int_0^1 \int_0^\infty \alpha \frac{\sqrt{1-\beta^2}}{\sqrt{1-\alpha^2}} \left (1-\exp\left[\frac{-2\tau_0\alpha}{\sqrt{1-\beta^2}} e^{-x^2} \right] \right) dxd\alpha d\beta}
\end{equation}

\end{tiny}

where once again $\tau_0$ is the optical depth of the 1038 \AA\ transition at line center, but now the
pathlength {\em l} is the 
diameter of the cylinder, while $\mu^\prime$ = $cos \theta$ where $\theta$ is the angle between the perpendicular to the curved surface of the cylinder and the observer's line-of-sight (see Figure 2).

Both of the above equations have been evaluated using numerical integration, and in Figures 3 and 4 we show
theoretical I(1032)/I(1038) intensity ratios for the slab and cylindrical geometries, respectively, for several different values of $\theta$. However, rather than plot results as a function of $\tau_0$, we instead use column density n$_e${\em l} (i.e. electron density $\times$ plasma thickness), as this is a commonly-employed 
quantity in astrophysical plasmas, and is reasonably straightforward to at least approximately estimate (see \S\ 4).
To derive the relationship between $\tau_0$ and n$_e${\em l}, we adopted a temperature of T$_{e}$ =
10$^{5.5}$ K for the formation of \ion{O}{6} in ionization equilibrium (Bryans et al. 2009), ratio of hydrogen to electron density of n$_H$/n$_e$ = 0.83 (i.e. that for a fully ionised plasma), oxygen to hydrogen abundance ratio of
 n$_O$/n$_H$ = 5$\times$10$^{-4}$ (Scott et al. 2009), oscillator strength f(1038) = 0.0673 (Aggarwal \& Keenan 2004), and (as in 
the rest of the analysis) the assumption of a Doppler line shape and identical electron and ion temperatures. 
This hence gave $\tau_0$(1038) = 2.392$\times$10$^{-18}$$n_e${\em l} (Rose et al. 2008). 
However, we stress that the relationship is approximate and depends on many assumptions, including for example the adoption of a single temperature for \ion{O}{6} formation when in fact it will be found over a range of values. Nevertheless, we believe plotting the figures using n$_{e}${\em l} as the x-axis rather than $\tau_0$ will be of most use to readers, who can if desired convert to $\tau_0$ using the relationship above.

We note that the I(1032)/I(1038) ratio tends to 2.0 for all values of $\theta$ at low and high column density in Figures 3 and 4
because the effect we are investigating depends on the ratio of escape factors between two lines (see equation 1), 
which only significantly differs from unity around optical depths of 1.

\section{Results and discussion}

We first note from Figure 1 that the ratios measured above the limb of the solar disk rapidly become much larger 
 than the value of 2.0 predicted for an optically thin plasma in coronal steady-state. However, it is well established that the \ion{O}{6} ratio can show very large values (up to $\sim$\,4) in spectra obtained for solar regions that lie above the surface, where the electron density is low (see, for example, Nakagawa 2008). As a consequence, processes other than collisional excitation can make a significant contribution to the \ion{O}{6} line emission, including the resonant scattering of chromospheric \ion{O}{6} radiation and/or the absorption and subsequent re-emission of Doppler-shifted \ion{C}{2} 1036.3 and 1037.0 \AA\ photons by \ion{O}{6} 1038 \AA\ (Kohl \& Withbroe 1982). Hence we do not consider further any observations made above the solar limb, and focus solely on those on the disk, where the \ion{O}{6} line emission will be dominated by the high electron density collisional excitation component, and therefore should have an optically thin I(1032)/I(1038) intensity ratio of 2.0. 
 
 For the ratio measurements on the solar disk, several show values significantly greater than 2.0, even allowing for observational uncertainties, 
  supporting the detection of line intensity enhancement due to opacity in our dataset. 
 Further support comes from a longitudinal analysis of the ratios in Figure 1. In such an analysis, we investigate if the changes in the observed ratios from one  position to the next within a region are correlated --- i.e. we are witnessing real changes in the ratio due to a variation in position --- or are they simply random  
 in nature. One would expect some correlation between ratio value and position, as the angle of observation $\theta$ and/or column density changes while moving from one location to another (Figures 3 and 4). 
 
The longitudinal analysis utilised a linear mixed effects model (Laird \& Ware 1982) to 
investigate the variation of line ratio with position within each region. We found that the within region covariance was highly significant, with a p-value of 
 $<$\,0.0001. The p-value is the probability that, within a region, the correlation of line ratios is zero. Hence our analysis indicates that the probability of this is $<$\,0.01\%, providing strong evidence to support the existence of such a correlation between the line ratios.
   
 A comparison of the observed ratios in Figure 1 with the theoretical results in Figures 3 and 4 provides some useful information on the emitting plasma. We note that the largest observed ratio is R = 2.24$\pm$0.02, in exact agreement with the maximum theoretical value for the slab geometry in Figure 3 of R = 2.24, providing observational support for the accuracy of the calculations. However, it also indicates that the emitting plasma must be slab-like in nature, as the maximum theoretical ratio for the cylindrical geometry in Figure 4 is only R = 2.10. Such a slab geometry might be expected for \ion{O}{6} emission, which arises from the upper transition region, being formed at a temperature of 10$^{5.5}$ K in ionization equilibrium (Bryans et al. 2009).
 
The smallest measured ratio, R = 1.51$\pm$0.01 near the solar limb, 
is also in good agreement with that predicted for the slab geometry in Figure 3, namely R = 1.48, while for the cylinder the lowest theoretical value is R = 1.41. Once again, this not only provides observational support for the theory but is consistent with what would be expected for a slab-like emission layer. At the limb, the line-of-sight to the plasma would be at a large angle $\theta$ to the perpendicular to a slab structure near the solar surface. For such large angles, the I(1032)/I(1038) intensity ratios are predicted to have their smallest values (Figure 3). By contrast, for a cylindrical geometry the line-of-sight could be at any angle to the cylinder surface (Figure 4), depending on the orientation of the cylinder to the observer. 

Both the largest and smallest ratio values in Figure 3 are predicted to occur at column densities n$_{e}${\em l} (where n$_{e}$ is electron density and {\em l} is pathlength) of around 10$^{17}$--10$^{17.5}$ cm$^{-2}$. Doschek et al. (1998) derived quiet Sun electron densities of n$_{e}$ $\simeq$ 10$^{9.7}$ cm$^{-3}$ from \ion{O}{5} diagnostic emission lines, which are formed at T$_{e}$ = 10$^{5.4}$ K (Bryans et al. 2009), similar to that for \ion{O}{6} (10$^{5.5}$ K). Hence the \ion{O}{5} density should reflect that in the \ion{O}{6} emitting region, which the solar atmospheric model of Avrett \&\ Loeser (2008) indicates has a thickness of around 500 km. Combining these plasma parameters yields a column density for the \ion{O}{6} region of n$_{e}${\em l} $\simeq$ 10$^{17.4}$ cm$^{-2}$, consistent with that predicted from Figure 3 to achieve values of the I(1032)/I(1038) ratio both significantly larger and smaller than the optically thin result of 2.0. However, we stress that our models do not assume constant n$_{e}$ nor {\em l}, with the theoretical
line ratios only dependent on the product n$_{e}${\em l} (or more precisely $\tau_0$), and hence these parameters can (and indeed very likely do) vary between observations.

A referee notes that calculations which include full geometry-dependent radiative transfer in realistic solar surface models (Wood \& Raymond 2000) indicate that the brightness of a low density region adjacent to a bright one can be significantly enhanced by scattered photons. If so, then one would expect an anticorrelation between values of I(1032)/I(1038) and the ratio of the local \ion{O}{6} intensity to that of the surrounding region, as positions with small 
local/surrounding intensity ratios (i.e. low brightness region adjacent to a brighter one) will show increased scattering and hence larger I(1032)/I(1038) ratios. However, comparisons of I(1032)/I(1038) measurements with brightness ratios derived using 7 $\times$ 7 pixel and 7 $\times$ 20 pixel areas reveal no such anticorrelation, indicating that the scattering process is unlikely to be responsible for the observed \ion{O}{6} line enhancements. We also note that if another plasma process were responsible for the apparent line enhancement effect, then the excellent agreement  between the present theory and SUMER measurements would be a major coincidence. 
   
In summary, we have (to our knowledge) provided the first definitive detection of intensity enhancement in solar transition region emission lines due to the effects of opacity. This detection not only confirms the presence of an interesting plasma process, but also illustrates how such observations can provide information on the physical conditions and geometry of the emitting plasma. The dependence of the line enhancement on plasma geometry is particularly important, as it may provide a way to diagnose, at least to some extent, the shape of an astrophysical source, most of which are spatially unresolved. For the future, it would be interesting to extend the work to obtain time-series spectra of a solar feature (e.g. active region) as it moves across the solar surface from the limb to disk center due to rotation, hence changing its orientation with respect to an observer and allowing a 3-dimensional map of the feature to be developed for comparison with theory. Additional observations of stellar and other potential time-varying astrophysical sources, once again obtained over significant amounts of time (e.g. a stellar rotation period), would also be of use to assess what geometrical information on the emitting plasma could be inferred from the line enhancement technique. 

On the theoretical side, we note that our opacity calculations assume a static solar atmosphere, when in reality it is highly dynamic. Hence, as noted by a referee, it would be useful if opacity calculations for \ion{O}{6} could be incorporated into a complex stellar atmosphere simulation code such as Bifrost (Gudiksen et al. 2011), which would allow a more realistic comparison of theoretical results with (dynamic) solar and stellar observations. In terms of our own theoretical work, 
our next steps will include more detail of surface variation and also the possibility of scattering involving different regions of the plasma.



\acknowledgments

FPK, JGD, and MM are grateful to the Science and Technology Facilities Council of the UK for financial support. MSM acknowledges financial support from the Leverhulme Trust.
Research at Armagh Observatory is
grant-aided by the N. Ireland Department of Culture, Arts and
Leisure (DCAL). SoHO is a mission of international co-operation between NASA and ESA.



{\it Facilities:} \facility{SOHO (EIT)}, \facility{SOHO (SUMER)}.

\begin{figure}[htp!]
\vspace{20cm}
 \includegraphics{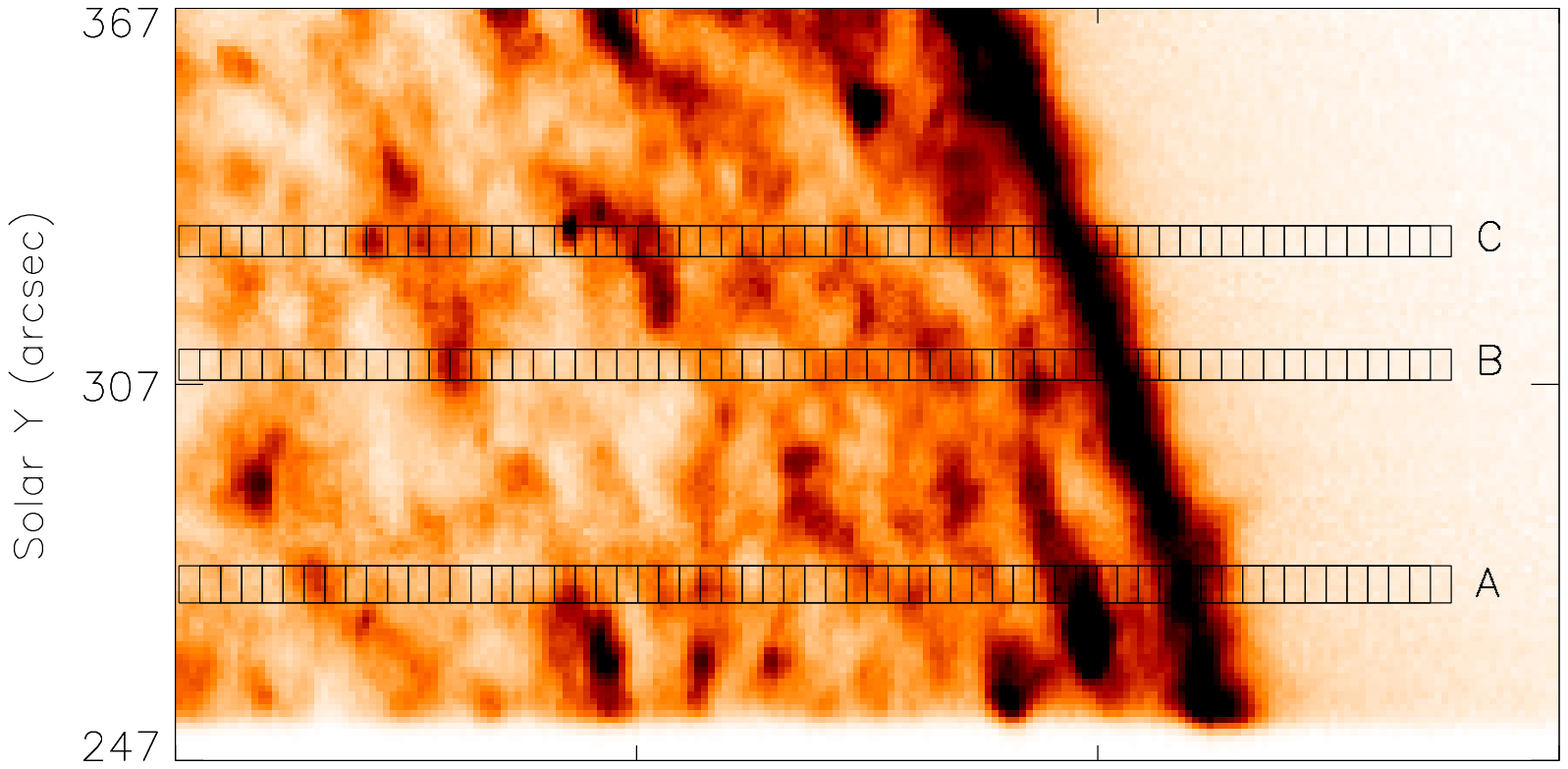}
 \includegraphics{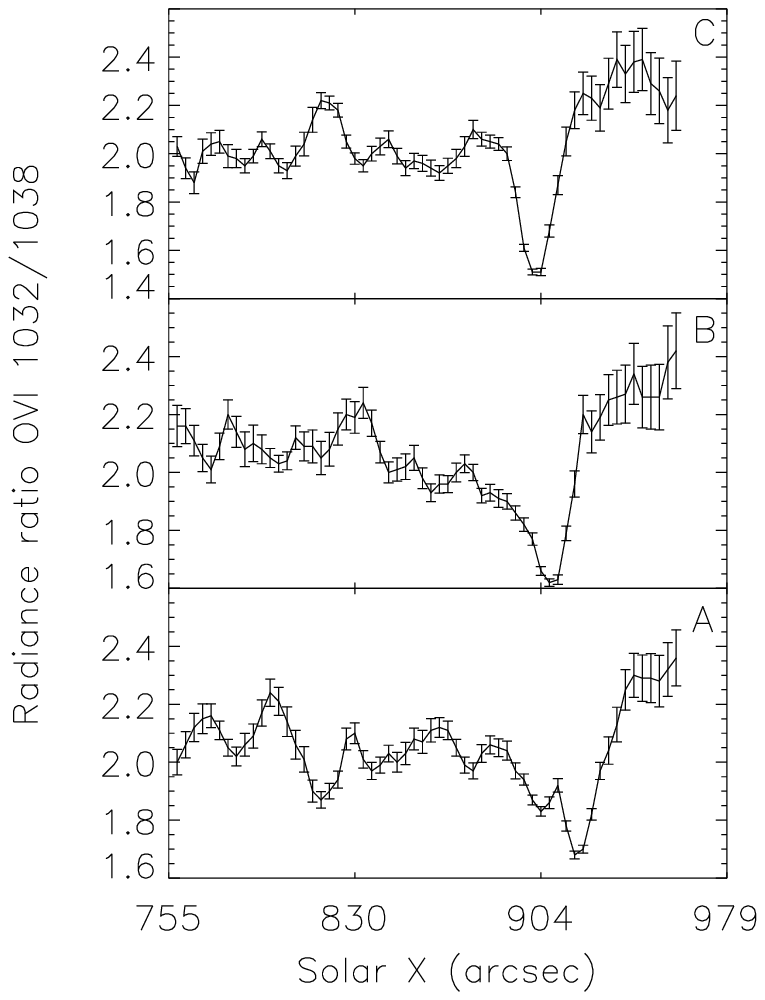}

\vspace*{0.1cm}

\caption{The top panel shows an image of the Sun in the \ion{O}{6} 1032 \AA\ line 
obtained on 1996 June 7 by the 
SUMER instrument, with the three regions for which \ion{O}{6} 1032 and 1038 \AA\ line intensities were determined from SUMER spectra 
labelled as A, B and C. The remaining panels show plots of the measured I(1032)/I(1038) ratios 
 for each region.
}
\end{figure}

\begin{figure}
\includegraphics[scale=0.8,angle=0]{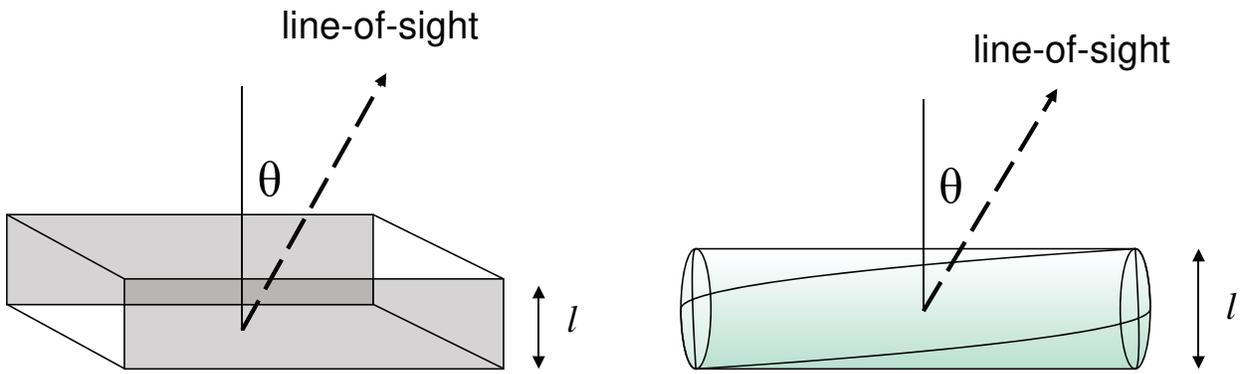}
\caption{Schematic diagram of the infinite plane slab (left-hand panel) and infinite cylindrical (right-hand panel) geometries discussed in \S\ 3. The line-of-sight to the observer is at an angle $\theta$ to the perpendicular to the slab or curved surface of the cylinder, both of which are of thickness $l$. 
}
\end{figure}

\begin{figure}
\includegraphics[scale=0.45,angle=0]{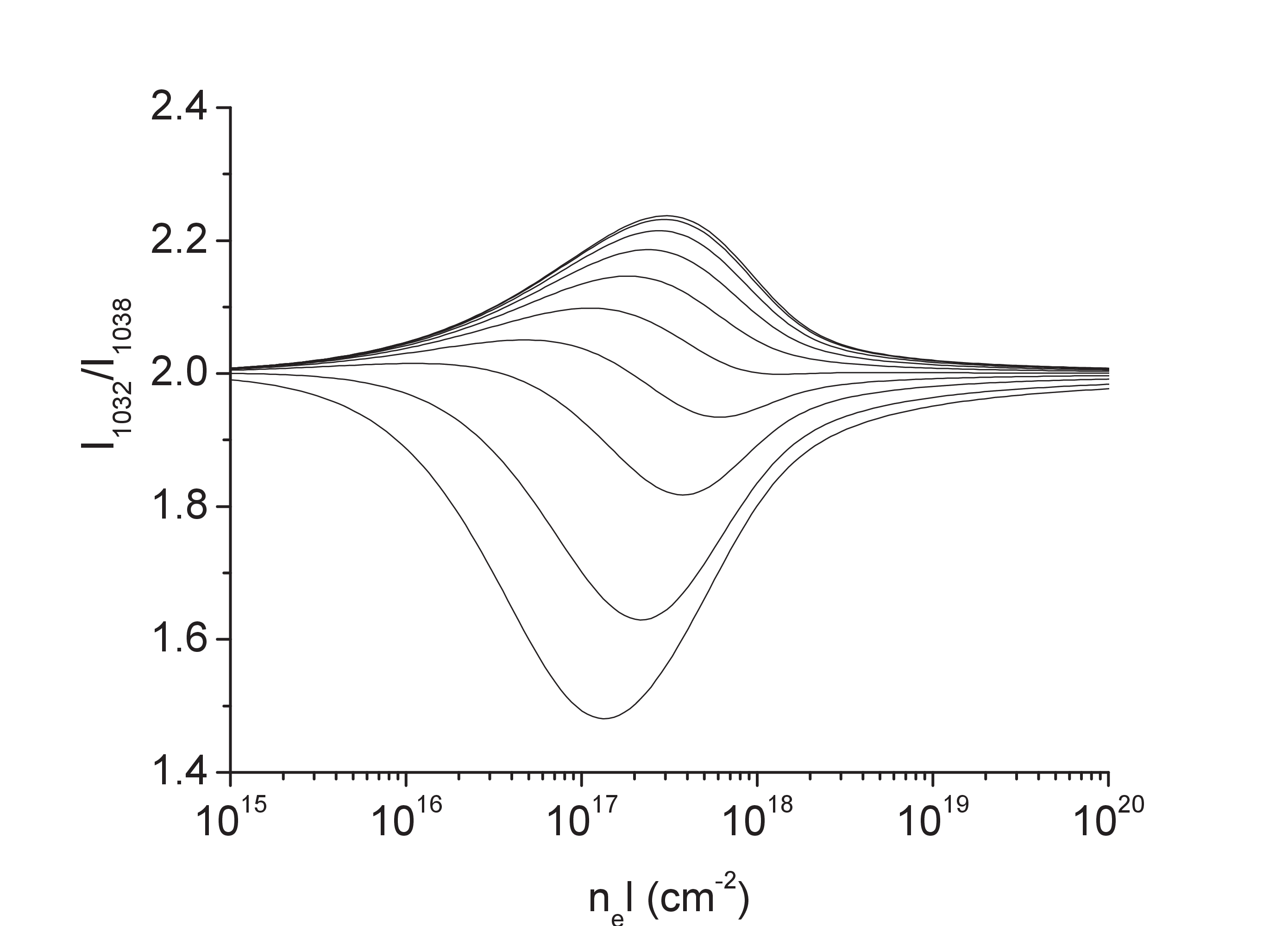}
\caption{Plots of the theoretical \ion{O}{6} I(1032)/I(1038) emission line intensity ratio as a function of column density $n_e l$ for the infinite plane slab plasma illustrated in Figure 2, for angles of observation (from top to bottom) of $\theta$ = 0, 10$^\circ$, 20$^\circ$, 30$^\circ$, 40$^\circ$, 50$^\circ$, 60$^\circ$, 70$^\circ$, 80$^\circ$
and 85$^\circ$. 
}
\end{figure}

\begin{figure}
\includegraphics[scale=0.45,angle=0]{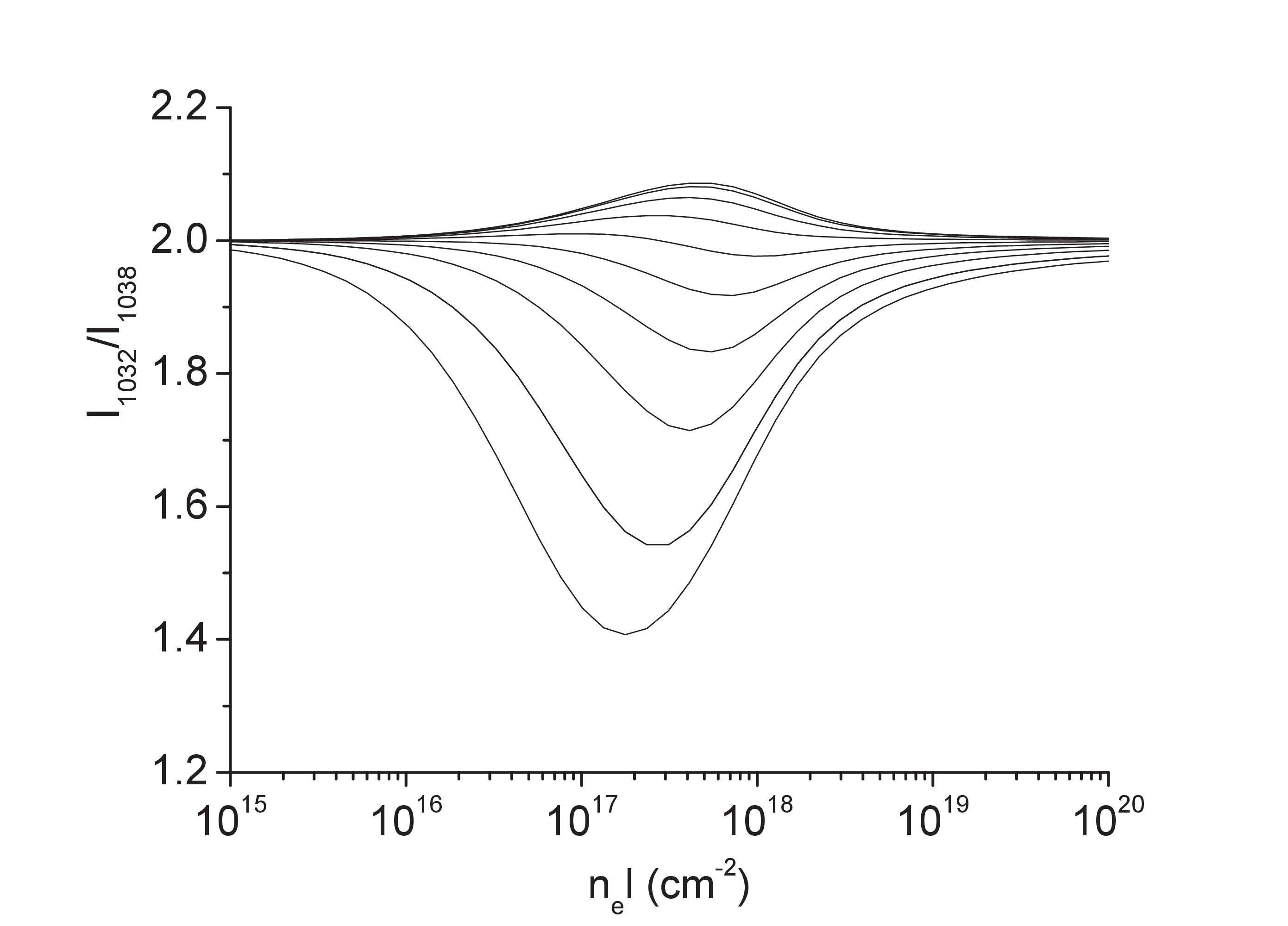}
\caption{Same as Figure 3 except for the infinite cylindrical plasma illustrated in Figure 2. 
}
\end{figure}

\end{document}